\documentclass[useAMS,usenatbib]{mn2e}%\documentclass{mn2e}
\usepackage{epsfig}
\usepackage{epsf}
\usepackage{graphicx}
\usepackage{natbib}
\usepackage{amsmath}\usepackage{amssymb}\usepackage{stfloats}

\title[Subdivided samples of LBVs]{The isolation of Luminous Blue
  Variables: On subdividing the sample}

\author[Smith]{Nathan Smith$^1$\thanks{Email: nathans@as.arizona.edu}
  \\ $^1$Steward Observatory, University of Arizona, 933 North Cherry
  Avenue, Tucson, AZ 85721, USA}

\begin{document}
\date{Accepted 0000, Received 0000, in original form 0000}
\pagerange{\pageref{firstpage}--\pageref{lastpage}} \pubyear{2002}
\def\arcdeg{\degr}
\maketitle
\label{firstpage}

\begin{abstract}

  A debate has arisen concerning the fundamental nature of luminous
  blue variables (LBVs) and their role in stellar evolution.  While
  Smith \& Tombleson proposed that their isolated environments
  indicate that LBVs must be largely the product of binary evolution,
  Humphreys et al.\ have recently expressed the view that the
  traditional single-star view still holds if one appropriately
  selects a subsample of LBVs.  This paper finds the claim of
  Humphreys et al.\ to be quantitatively unjustified.  
%% this is a nice way to say it.
  A statistical test of ``candidate'' as opposed to ``confirmed'' LBVs
  shows no significant difference ($<$1$\sigma$) between their
  environments.  Even if the sample is further subdivided as proposed,
  the three most luminous LBVs are spatially dispersed similar to late
  O-type dwarfs, which have much longer median lifetimes than expected
  for classical LBVs.  The lower-luminosity LBVs have a distribution
  associated with red supergiants (RSGs), but these RSGs are dominated
  by stars of 10-15 $M_{\odot}$ initial mass, with much longer
  lifetimes than expected for those lower-luminosity LBVs.  If one's
  view is restricted to the {\it highest-luminosity} LBVs, then the
  appropriate comparison is with {\it early} O-type stars that are
  their presumed progenitors; when this is done, it is clear that even
  the high-luminosity LBVs are more dispersed than expected.
  Humphreys et al.\ also suggest that velocities of LBVs support the
  single-star view, being inconsistent with runaways.  A quantitative
  analysis of the radial velocity distribution of LBVS in M31 and M33
  contradicts this; modest runway speeds expected from mass gainers in
  binary evolution are consistent with the observed velocities,
  although the data lack the precision to discriminate.

\end{abstract}


%%%%%%%%%%%%%%%%%%%%%%%%%%%%%%%%%%%%%%%%%%%%%%%%%%%%%%%%%%%%%%%%%%%%%%
\section{Introduction}

The eruptive mass loss exhibited by luminous blue variables (LBVs) is
thought to be an important ingredient in stellar evolution (see, e.g.,
\citealt{so06,smith14}), but the way LBVs actually figure in this
evolution and the physical mechanisms of their outbursts remains
challenging to understand.  Moreover, LBVs are thought to be related
to some extragalactic non-supernova (SN) transients
\citep{smith11,vdm12} and their mass loss is reminiscent of extreme
pre-SN eruptions \citep{smith14}.

In a recent study, Smith \& Tombleson (2015; ST15 hereafter) analyzed
the projected spatial distribution of LBVs on the sky and found them
to be surprisingly isolated from O-type stars. ST15 concluded that the
results were inconsistent with expectations for the traditional
picture of LBVs in single-star evolution (e.g., \citealt{hd94}),
wherein LBVs are descended from very massive main sequence O-type
stars, and where LBVs are the key agent that provides the required
mass loss to drive them into the Wolf-Rayet (WR) phase.  In
particular, ST15 found that LBVs were more dispersed from O stars on
the sky than WR stars, making it impossible for the observed
population of LBVs to turn into the observed population of WR stars.
ST15 concluded that many LBVs are likely to be the product of binary
evolution, where stars are spun-up, chemically enriched, and
rejuvinated by mass transfer, and possibly kicked by their companion's
SN explosion.  In this view, LBVs are evolved massive blue stragglers.

Humphreys et al. (2016; H16 hereafter) present a contrasting
viewpoint, arguing that environments of LBVs are instead consistent
with the traditional view if one divides and culls the sample of LBVs
in the way they prefer.  They also claim that observed kinematics
indicate that none of the confirmed LBVs are runaway stars.  The
discussion below critically examines these claims, since the role
played by LBVs and their mass loss is fundamental to our understanding
of massive star evolution and the origin of WR stars.  Essentially, it
is found that the claims made by H16 are not quantitatively
justifiable based on the data, even if one permits the selective
subdivision of the LBV sample as they envision.  In some cases the
quantitative implication of the data yields the opposite of their
qualitative interpretation.

This paper undertakes a critique of the claims made by H16.  Before
examining their analysis, Section 2 first corrects some errors and
points out misconceptions that influence the data and expectations in
the H16 paper.  Section 3 concentrates on the claimed difference
between confirmed and candidate LBVs (each corresponding to about half
the sample in ST15), showing that their environments are
indistinguishable with available data.  Section 4 reasseses the
analysis of H16; here we permit the subjective subdivision of the
sample as preferred by H16, and we ignore small number statistics.
Even with these accomodations, we find that the central conclusions
found by H16 were not the correct conclusions indicated by their
suggested division of the data.  This is because if one wishes to
split a cumulative distribution, one must also split the distribution
of comparison objects in order to draw a meaningful conclusion.
Finally, Section 5 provides a quantitative analysis of the claims by
H16 regarding radial velocities, finding them to be invalid.  Runaway
LBVs are certainly allowed by the radial velocity data, although not
clearly required.

\section{Some corrections and misconceptions}

Although \citet{conti84} originally defined LBVs rather broadly to
include a number of types of luminous blue supergiants (BSGs) with
high mass loss, H16 provide explanations for why they choose to
include or exclude certain stars from ST15's statistical sample of
LBVs in the Magellanic Clouds.  This is discussed below.  Even if
those qualifications are accepted, there are some factual errors and
misconceptions in the H16 paper that should be corrected before one
considers their interpretation.

An apparent error concerns HD~5980, the most luminous star in the SMC.
H16 note that numerous authors describe HD~5980 as a confirmed LBV
that suffered a giant eruption, and they therefore choose to include
it in their Table 2 as an LBV.  However, they do not include HD~5980
in the cumulative distribution plot of separations from O stars (their
Fig.\ 5), even though they do include the lower-luminosity star R40 in
the SMC.  This might significantly influence the main claim of their
paper, since with such a high luminosity, HD~5980 should be included
with the high-luminosity LBVs (i.e. the ``LBV1'' group), of which
there are only three objects.  Adding a fourth, which happens to have
the highest separation from O stars ($\sim$21 pc) among this sample,
skews the distribution a noticable amount.  H16 do not state a reason
for omission of this star from the cumulative distribution, so one may
assume this was an oversight.

\begin{table}
\begin{center}\begin{minipage}{3.3in}
    \caption{Coordinates for two stars claimed to be the same star by
      Humphreys et al.}\scriptsize
\begin{tabular}{@{}lccl}\hline\hline
Sanduleak name  &R.A.(J2000) &DEC.(J2000)  &other name \\ \hline
Sk $-$69$^{\circ}$142a &05:27:52.657  &$-$68:59:08.56   &HDE~269582  \\
Sk $-$69$^{\circ}$147  &05:28:21.987  &$-$68:59:48.20   &MWC~112    \\
\hline
\end{tabular}\label{tab:sk}
\end{minipage}\end{center}
%$^a$A.
\end{table}

H16 claim that ST15 have a duplicate entry in their sample of LBVs,
but this is false.  Referring to ST15's sample, H16 state that
``MWC~112 was included twice; it is the same as Sk
$-$69$^{\circ}$142a''.  However, consulting the SIMBAD database
reveals that MWC~112 is in fact not $-$69$^{\circ}$142a as they claim;
the correct Sanduleak designation of MWC~112 is Sk $-$69$^{\circ}$147.
The J2000 coordinates of the two stars, which are 2$\farcm$82 apart on
the sky, are reproduced in Table~\ref{tab:sk}.  MWC~112 is listed as
an F5~Ia supergiant in the LMC (not WN10h as listed in Table 3 of H16;
this should be for Sk $-$69$^{\circ}$142a).  Whether or not this is
the same star that was concluded to be an LBV by \citet{vGS96} is
unclear, since some authors have confused MWC~112 and Sk
$-$69$^{\circ}$142a multiple times before in the literature (see, for
example, \citealt{hd94} and \citealt{vGS96}, where in both cases
MWC~112 is erroneously listed as the same star as HDE~269582 = Sk
$-$69$^{\circ}$142a).  Interestingly, in the original Mount Wilson
Catalog \citep{merrill33}, MWC~112 is listed as a type ``Beq'' with no
luminosity class (P Cygni is listed as B1eq in the same catalog) with
a photographic magnitude of 13.0 mag, whereas \citet{rousseau78} list
it as F5~Ia with $B$=11.8 mag.  This seems to suggest a change in
spectral type and magnitude reminiscent of LBVs that are true S
Doradus variables (i.e. 1.2 mag brighter in its cooler eruptive
phase), although a renewed examination and perhaps monitoring to
resolve this would be worthwhile.  This correction is important for
future studies, but has little influence on the results here because
MWC~112 is roughly in the middle of the distribution of O-star
separations.

LBVs reside in a part of the HR Diagram that overlaps with other stars
of similar luminosity ($L$) and apparent temperature ($T$) that have
not been observed to exhibit the same instability.  H16 state that
LBVs are known to be distinguished from these other stars with similar
$L$ and $T$ because they have higher $L$/$M$ ratios.  However, this is
an assertion that has no empirical verification.  There is no LBV with
a stellar mass that has been measured directly in a binary system; the
idea that they have higher $L$/$M$ ratios than other stars of similar
$L$ is a conjecture from single-star evolutionary models, not
observations.

There is another misconception expressed by H16, which is key to one
of their central qualitative arguments.  They claim that the observed
radial velocities of LBVs in M31/M33 and the LMC are consistent with
their positions in those galaxies, and that except in one case, there
is no evidence for nebular bowshock morphology that is expected if the
stars are runaways.  They cite these points as a counterargument that
LBVs are not likely to be runaway mass gainers.  H16 do not quantify
the velocities or velocity distribution that they expect, but based on
the arguments and information in the paper, one might conjecture that
they expect runaway velocities in excess of 100 km s$^{-1}$.  This is
based on their claim that observed LBV velocities seem to be within
$\pm$40 km s$^{-1}$ of the expected velocities from rotation curves
(and are therefore consistent with no runaways), and also the claim
that such motion would significantly influence the morphology of
nebulae that are expanding at 10--50 km s$^{-1}$.  H16 assert that
these motions are consistent with not being runaways, although they
give no analysis to support this.  This point is addressed below.
Here the misconception about the expected velocities is clarified.

In a close binary that has experienced mass transfer, with one star
being the mass gainer, the mass donor will likely explode first
(although not always first) as a stripped-envelope SN and may provide
a kick to the mass gainer.  This is one of the ideas to help explain
the distribution of LBVs (ST15), which H16 argue against.  Here, the
``kick'' might be recoil from a neutron star kick and an asymmetric
SN, which could induce a fairly high speed runaway in the companion.
However, even without this explosive kick, a more likely case is that
the mass gainer will have resulting motion that is simply its orbital
velocity and trajectory at the time when its companion explodes (note
that the compact companion may also remain bound and the binary system
may have net motion).  The key point here is that the mass gainer has
become much more massive than its companion (which is now a much
lower-mass He core that has donated its H envelope to the mass
gainer).  As a consequence, the mass gainer's orbital speed will not
be so fast at the time of explosion.  For a 30 $M_{\odot}$ mass gainer
and a $\sim$4 $M_{\odot}$ stripped He core mass donor, for example,
expected kick velocities are only of order 5-20 km s$^{-1}$ depending
on the orbital period appropriate for systems that do not
merge.\footnote{Somewhat higher runaway speeds can be achieved with
  lower-mass gainers or non-conservative mass transfer, but these
  wouldn't be the cases expected to yield an LBV-like object.}  Models
of binary stellar populations predict relatively slow runaways --- or
``walkaways'' --- of only a few to 10 km s$^{-1}$ in many cases
involving mass stripping \citep{eldridge11,demink14}.  Faster runaways
are actually hard to get from binary evolution without a favorable
neutron star kick, and may require a 2-step ejection
\citep{pk10}. Therefore, H16 were not correct to conclude that the
observed velocities argue against the presence of runaways among LBVs;
the radial velocity values they listed are unable to quantitatively
justify such a claim (see below).  It is not at all clear that modest
speeds of this order would lead to easily identified bow shock
morphologies, as asserted by H16, because the runaway speed may
actually be substantially less than the speed of the nebular shell.
However, even such modest speeds of $\sim$10 km s$^{-1}$ are enough to
move LBVs a few 10s of pc in a few Myr.  A key point is that the
relatively slow kick velocity acts together with a longer lifetime
(longer than expected in single-star evolution for the observed
luminosity) to make the stars appear more isolated.  The longer
lifetime allows O-type stars from the same birth population to die in
the mean time.

In a similar vein, H16 argue that simple OB association random drift
velocities of 10 km s$^{-1}$ can yield the inferred motions of LBVs
required to explain their degree of isolation.  However, the critical
point here is that random velocity dispersion in a cluster or OB
association applies to {\it all} stars, whereas the projected
distribution of LBVs on the sky indicates that they are {\it
  preferentially} more isolated than even the late-type O stars.  In
other words, statistically LBVs appear to receive an {\it extra}
velocity spread (or longer lifetimes than expected), beyond that given
to the rest of massive stars.  This is why one must consider an
appropriate comparison between a full {\it distribution} of
separations, rather then picking a few stars out of a sample to
conclude that a late O-type star or RSG is seen nearby.  H16 noted
some projected massive star neighbors to LBVs in M31/M33 and deemed
this sufficient to claim that they are consistent with single-star
evolution, but they did not present an appropriate statistical
analysis.

Regarding binarity, H16 stated that some LBVs are observed to be in
binary systems, and that this therefore contradicts the scenario
proposed by ST15 where LBVs are mass gainers.  This statement is false
for two reasons.  First, in the scenario where LBVs form through mass
transfer in binaries, there will indeed be some cases where the mass
donor companion has exploded, leaving a runaway single star, but there
will also be some cases where this has not yet occurred so that the
LBV is still in a binary system awaiting its first SN.  The time for
the mass gainer to evolve to the LBV phase and the time for the mass
donor to explode depend on various factors, including the initial
masses and how much mass was transferred (see, e.g.,
\citealt{langer12}).  It would, however, be interesting to investigate
the nature of these companions of LBVs to see if they are consistent
with being a mass donor.  Second, even if LBVs are the products of
mergers or if their mass donor has already exploded, LBVs can
potentially still be seen in binary systems (especially wide binaries)
with unevolved companions because hierarchical triple and multiple
systems are the norm among massive stars, not the exception
\citep{abt90,sana12,dk13,sana14,rizzuto13,moe16}.  These studies
generally find companion frequencies per primary around 2 (i.e. triple
systems are very common).  Combining unresolved close
spectroscopic/eclipsing binaries with spatially resolved wide
companions, \citet{sana14} estimate that the frequency of multiple
systems may well be 100 per cent among O stars in particular.
Furthermore, with the relatively weak SN kicks expected in the
mass-gainer scenario, the resulting binaries (which were originally
triples) may survive disruption and may yield highly eccentric and
wide orbits.  Interestingly, the companions that have been noted for
LBVs so far are all in wide orbits: $\eta$ Car has a 5.5 yr orbital
period with very high eccentricity \citep{damineli97}; HR~Car is a
wide resolved binary with a period close to 5 yr \citep{rivinius15};
both HD~168625 and MWC314 have wide companions that are directly
imaged \citep{martayan16}.  The interesting question to ask is not
whether there are some LBVs in binaries, as noted by H16, but rather,
whether there is a statistically significant difference in the binary
fraction of LBVs as compared to that of O stars and other types of
evolved massive stars. \citet{martayan16} estimate a binary fraction
among LBVs of only 27 per cent, whereas it is much higher for O-type
stars (essentially 100 per cent if one includes wide binaries, as
noted above).  This fraction is admittedly preliminary, since fainter
companions are hard to detect around very bright LBVs.

%%%%%%%%%%%%%%%%%%%%%%%%%%%%%%%%%%%%%%%%%%%%%%%%%%%%%%%%%%%%%%%%%%%%%%%
\begin{figure}\begin{center}
    \includegraphics[width=2.9in]{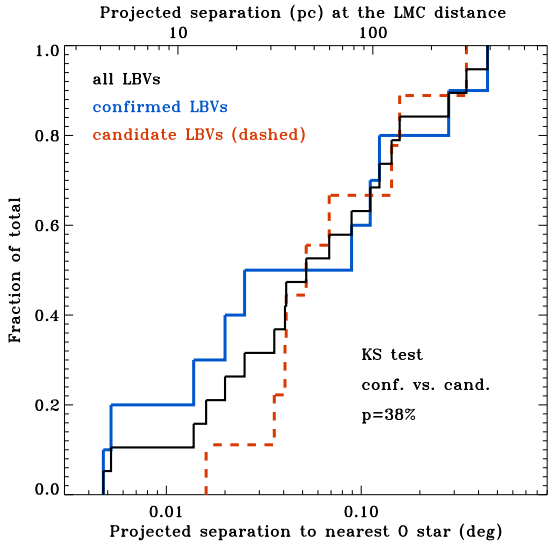}
\end{center}
\caption{Cumulative distributions of separations from O stars for LBVs
  from ST15, plotting all LBVs together (black) as well as separate
  distributions for confirmed (blue) and candidate (dashed orange)
  LBVs.  The result of a KS test between the confirmed and candidate
  distributions is noted.}
\label{fig:ks}
\end{figure}
%%%%%%%%%%%%%%%%%%%%%%%%%%%%%%%%%%%%%%%%%%%%%%%%%%%%%%%%%%%%%%%%%%%%%%%

\section{Confirmed vs. candidate LBVs}

LBVs are rare because massive stars are intrinsically rare, because
they are in a brief evolutionary phase, because their variability
seems to turn on and off, and perhaps also because they are the
product of special circumstances.  When faced with these small
numbers, selecting out only the top end of the subset that meets some
specific criteria leaves one with a sample size that is too small to
make any statistical claims.  An interesting question is whether or
not there is reason to separate LBVs confirmed to have exhibited the
characteristic variability from those that resemble them in terms of
their quiescent properties (candidates), which is about half-and-half.

H16 proposed that lumping together confirmed and candidate LBVs led
ST15 to an incorrect conclusion about the statistics of their
environments.  A KS test allows one to determine if such a claim is
quantitatively valid.  Figure~\ref{fig:ks} shows the distributions of
projected separation to the nearest O-type stars (D1 from Table 1 of
ST15, for Magellanic Cloud LBVs only), plotted with all LBVs lumped
together (black) and with the confirmed LBVs (blue) separated from the
candidate LBVs (dashed orange).  A KS test of these two separate
distributions gives a p-value of 38 per cent (less than 1$\sigma$).
To claim that confirmed and candidate LBV populations are drawn from
separate parent distributions, a p-value less than 5 per cent is
required.  Thus, confirmed LBVs and LBV candidates are quantitatively
consistent with being drawn from the same sample as far as their
environments are concerned.  The claim by H16 that not separating them
will corrupt the statistical interpretation of their environments is
therefore found to be invalid.

Confirmed LBVs do seem to have a tail reaching to smaller
distributions than the candidates in Figure~\ref{fig:ks}, which was a
central point made by H16.  There is no statistical significance to
this difference, but there may be some physical significance if this
were to persist in a larger sample size.  In that case, one might
simply infer that the most luminous LBVs have shorter duty cycles in
their eruptive behavior (perhaps because of a closer proximity to the
Eddington limit), making it more likely for them to be confirmed by
their observed variability on timescales of modern observations.
There is also a selection effect that the more luminous LBVs have a
larger change in temperature in an S Dor cycle, and hence a larger
amplitude in their visible brightening, which makes them easier to
detect.  Lower-luminosity LBVs may have longer periods between the
emergence of their eruptive variability and smaller amplitudes in
their brightening, making it more likely for them to be disregarded as
mere candidates, even if they originate from a similar evolutionary
pathway and the same distribution of environments.

%%%%%%%%%%%%%%%%%%%%%%%%%%%%%%%%%%%%%%%%%%%%%%%%%%%%%%%%%%%%%%%%%%%%%%%
\begin{figure*}\begin{center}
    \includegraphics[width=5.9in]{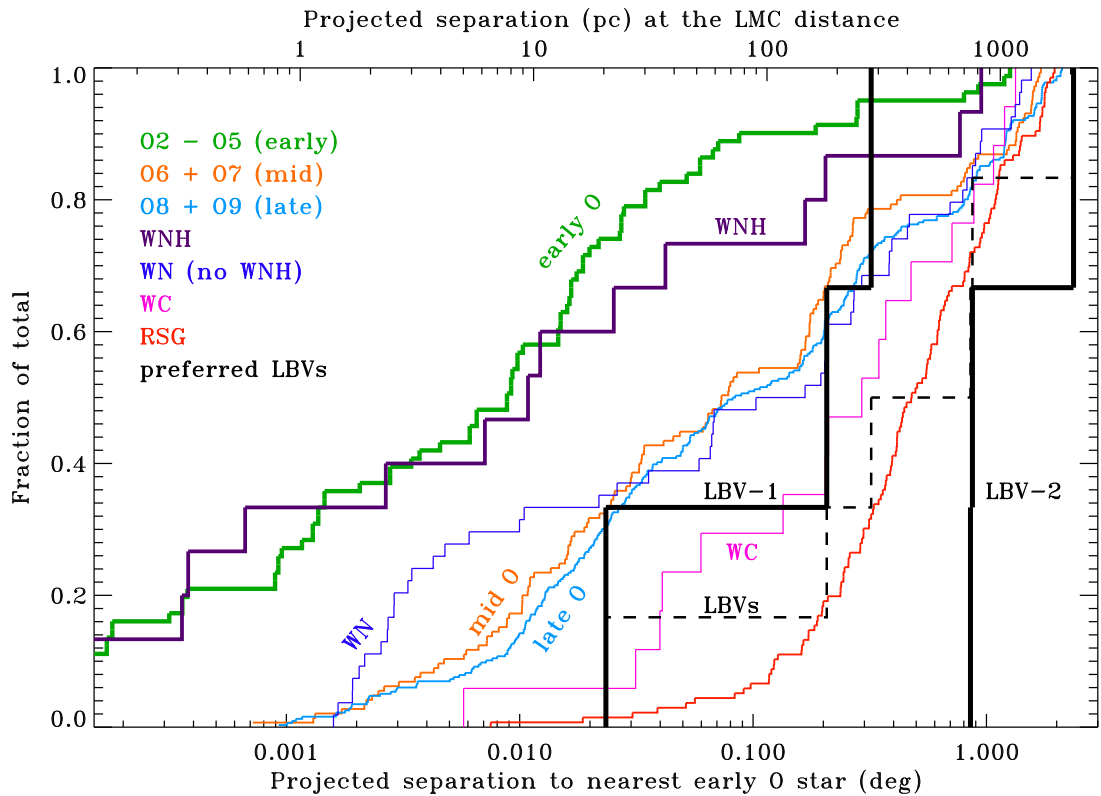}
\end{center}
\caption{This is similar to the cumulative distribution plot in Figure
  4 from ST15 (i.e. using the same star positions), except that (1) it
  shows the projected separation on the sky to the nearest {\it early}
  O-type dwarfs (O2-O5 V) rather than the separation to {\it any}
  O-type dwarf, and (2) it subdivides the confirmed LBVs into two
  groups as preferred by H16 (LBV1 and LBV2 are solid black lines, and
  the combination LBV1+LBV2 is the dashed black line.)  Also, WNH
  stars are plotted separately from other WN stars.}
\label{fig:cumplot}
\end{figure*}
%%%%%%%%%%%%%%%%%%%%%%%%%%%%%%%%%%%%%%%%%%%%%%%%%%%%%%%%%%%%%%%%%%%%%%%

\section{Cherry-picking the sample}

The central argument made in the original paper by ST15 was a
statistical one.  Namely, when one considers the {\it full
  distribution} of LBVs and related stars, they have a projected
separation on the sky that appears to be inconsistent with being
descended from the observed distribution of O-type stars, and
moreover, inconsistent with being the precursors of the observed
distribution of WR stars.  Considering only stars in the Magellanic
Clouds, ST15 made cumulative distribution plots of the projected
separations on the sky between various different types of stars and
their nearest O-type neighbors, finding that LBVs do not reside where
one expects.  ST15 therefore concluded that binary evolution is likely
to be important for explaining origin of LBVs.

H16 have recently offered a contrasting viewpoint, arguing that this
already small sample of stars should be further subdivided. They
suggested that ST15 confused high-luminosity classical LBVs with
lower-luminosity LBVs, which, as some have argued, do appear to fall
into two possibly separate groups \citep{smith04}.  When culling of
the sample is performed in the way they describe, H16 conclude that
LBVs are consistent with single-star evolution after all.  This
contradiction warrants a detailed examination.

Adopting criteria from \citet{hd94}, H16 take the sample of 19
Magellanic Cloud LBVs from ST15 and subdivide it into three smaller
groups: (1) confirmed classical LBVs with only three members (called
group `LBV1'), (2) low-luminosity confirmed LBVs with 4 members
('LBV2'), and (3) unconfirmed or `candidate' LBVs that are the
remainder.\footnote{Note that other authors have suggested somewhat
  different subdivisions of LBVs \citep{vg01}.}  With this subdivided
sample, H16 remake the cumulative distribution plot in Figure 4 from
ST15 and infer very different conclusions.  Namely, selecting only the
three objects that are confirmed classical LBVs at high luminosity --
which also happen to have the three smallest projected separations out
of the other 19 objects --- they find that the cumulative distribution
of the LBV1 separation is smaller than the full sample of LBVs from
ST15.  They take this as an indication that LBVs in general are not as
widely dispersed as claimed by ST15.  Morover, they find that when
only the LBV1 subsample is considered, that its spatial distribution
on the sky overlaps with late O-type dwarfs.  From this, H16 conclude
that the LBV1 group is consistent with traditional single-star views
of LBVs.  Similarly, they find that the lower-luminosity LBV2
subsample overlaps with the distribution of red supergiants (RSGs),
and H16 conclude again that this supports the traditional view,
wherein lower-luminosity LBVs are thought to be post-RSGs.

One may question the validity of selecting the tail end of a
distribution of objects and then remaking the cumulative distribution
comparison in the same way, not to mention the statistical
significance of only three objects in the LBV1 group with unquantified
selection bias.  Nevertheless, even if this sort of cherry-picking of
an already small sample is permitted, one can see that H16 interpreted
their results incorrectly.  To demonstrate this, let's separate the
two cases of LBV1 and LBV2 below.

For the case of the LBV1 group, H16 find that their distribution of
separations overlaps with late O-type stars, and they conclude that
this association supports the traditional single-star view. However,
if one extracts the tail end of a distribution, one must take care in
the resulting comparison.  These three stars are the high-luminosity
classical LBVs in the LMC (R127, R143, and S Dor), which are
representative of very luminous LBVs with implied initial masses (in
the single-star evolutionary picture) of something like 60-100
$M_{\odot}$ and lifetimes of $\sim$2.5--3 Myr.  H16 found that these
have a similar distribution on the sky to late O-type dwarfs (O8 V and
O9 V in the sample of ST15), which have initial masses of 18-22
$M_{\odot}$ and lifetimes of roughly 9-11 Myr.  H16 should therefore
have concluded from this comparison that classical LBVs are very
overluminous for their distribution on the sky.

For the lower-luminosity LBV2 sample, H16 find a dispersal on the sky
similar to RSGs and conclude that this is consistent with these stars
being post-RSGs as in the traditional view \citep{hd94}.  However,
here again one must be somewhat quantitative about what ages and
initial masses are actually implied by each subgroup.  The
low-luminosity LBVs have luminosities that would require initial
masses of 25-40 $M_{\odot}$ if they are evolved single stars (see,
e.g., \citealt{smith04}, and ST15).  If these are post RSGs, then they
should be associated with the very most extreme RSGs with similar
initial masses.  However, the RSG comparison sample from ST15 included
{\it all} the RSGs in the LMC.  As noted by ST15, by number this is
dominated by the low end of the mass distribution because of the slope
of the initial mass function.  Thus, the RSG comparison sample
represents mostly stars of $\sim$10-15 $M_{\odot}$ initial mass, with
ages of 20-100 Myr.  These RSGs were not O-type stars on the main
sequence; they were early B-type stars.  The fact that low-luminosity
LBVs are so well associated with them does {\it not} support the view
that these LBVs are the product of mass loss from the most extreme
RSGs --- in fact, it negates this possibility.  Again, from this
selective comparison, H16 should have concluded that the LBV2 group is
highly overluminous for its distribution on the sky, and therefore
very inconsistent with the traditional view of LBVs in single-star
evolution \citep{hd94}.  Unfortunately, serious confusion has arisen
about the distinction between garden variety RSGs of 10-15 $M_{\odot}$
and the more extreme RSGs like VY~CMa (see \citealt{smith09} for a
more detailed explanation).

If one wishes to extract only the tail end of a cumulative
distribution plot, one must then make an appropriate comparison in
order to draw the correct conclusion.  ST15 quantified separations
between various classes of massive stars and their O-type star
neighbors.  They did this by considering the nearest neighboring
O-type dwarf of {\it any} spectral subtype.  This was because ST15
were analyzing the full distribution of LBVs, which included both
higher- and lower-luminosity LBVs and candidates in order to produce a
statistically significant result.  If one wishes to extract only the
most luminous LBVs (as for the three objects in the LBV1 subsample
defined by H16), then it would be appropriate to compare
these to a subset of more massive and younger O-type stars within a
similarly restricted luminosity range.  In other words, only the early
O stars are the putative progenitors of the LBV1 group, not late
O-type stars, so early O stars should be the basis for comparison with
LBV1.

Figure~\ref{fig:cumplot} shows such a comparison.  This is essentially
the same as Figure 4 in ST15, but it plots the projected separation on
the sky to the nearest star in the LMC with an {\it early} O spectral
type (types O2-O5; the early O stars from ST15) rather than {\it any}
O-type dwarf.  This is a more appropriate comparison, since the early
O stars (main sequence masses above roughly 35-40 $M_{\odot}$) must be
the progenitors of classical LBVs in the traditional single-star view,
and since late O-type dwarfs are scattered throughout this region of
the LMC (as noted by ST15, many of the nearest neighbors to LBVs
appear to be accidental line-of-sight coincidence). Following the
suggestion of H16, LBVs are separated into the same LBV1
(classical) and LBV2 (low-luminosity) subgroups that they
prefer.\footnote{These groups are the same as H16 with
  one exception.  H16 chose to include the SMC star R40
  in the LBV2 group, but did not include the SMC star HD~5980 in the
  LBV1 group. Both are confirmed LBVs.  HD~5980 would have the largest
  separation from O stars on the sky and would have skewed the LBV1
  sample to even larger separations. Figure~\ref{fig:cumplot}
  therefore includes only LMC stars.}  This more appropriate
comparison with early O-type stars shows that even the exclusive
subsample of the most luminous LBVs (LBV1) is dispersed far from the
locations of their supposed progenitors.  LBV1 members are located
20-300 pc from the nearest early O stars, which reside mostly in young
clusters.  They are also more dispersed from O stars than most of the
WN stars, only the most luminous of which should be their descendants.
The main point in a plot like Figure~\ref{fig:cumplot} is not to claim
that any individual massive star can never be so far from other O
stars, but rather, that the distributions of LBV1 and early O stars
are inconsistent.

It is true that the LBV1 sample should be somewhat older than their
main sequence progenitors and that they should be slightly more
dispersed on the sky.  But how much more?  Figure~\ref{fig:cumplot}
shows that LBV1 has a distribution as old as or older than late O
dwarfs, with lifetimes of 9-11 Myr.  Since O stars typically spend the
first 0.5--1 Myr embedded in their natal clouds, and since O stars get
brighter as they age along the main sequence, the observed population
of late O stars has a likely median lifetime of at least 5--6 Myr, if
not older.  The early O stars have H-burning lifetimes of only 2.5-3
Myr and median lifetimes (spending their early childhood embedded in
ultracompact H~{\sc ii} regions inside molecular clouds) of roughly 2
Myr.  The high luminosity classical LBVs are thought to mark the
transition from the end of core-H burning to He burning in these very
massive stars (in the traditional picture), so they should have actual
ages of 2.5-3 Myr.  In fact, however, the LBV1 distribution has an
apparent age that is more than double this value (similar to or larger
than the median age of late O-type).  This suggests that their longer
lifetimes are more appropriate for a star of roughly {\it half their
  implied initial mass}.

%%%%%%%%%%%%%%%%%%%%%%%%%%%%%%%%%%%%%%%%%%%%%%%%%%%%%%%%%%%%%%%%%%%%%%%
\begin{figure*}\begin{center}
  \includegraphics[width=5.7in]{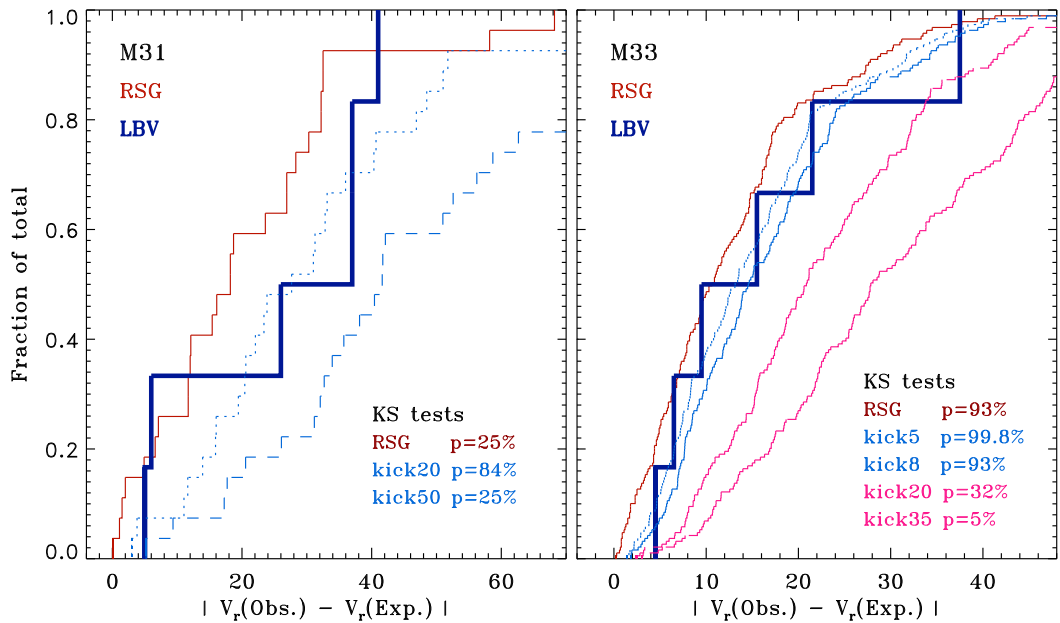}
\end{center}
\caption{Cumulative distributions of velocities for RSGs (red) and
  LBVs (dark blue) in M31 and M33.  Plotted here is the absolute value
  of the difference between the observed radial velocity $V_r(Obs.)$
  and the radial velocity expected from the rotation curve
  $V_r(Exp.)$.  Values for RSGs are ``rank 1'' (i.e. highly likely
  membership) sources from \citet{drout09} for M31 and \citet{drout12}
  for M33.  LBVs are calculated from values presented in Table 1 of
  H16.  Possible distributions with kicks are shown by
  adding a randomized kicks to the RSG distribution in each galaxy.
  Values shown are 20 and 50 km s$^{-1}$ for M31 and 5 and 8 km
  s$^{-1}$ for M33.  Listed in each panel are the results of KS tests
  comparing LBVs to the three other distributions. The additional two
  magenta curves for M33 show simulated distributions with kicks of 20
  and 35 km s$^{-1}$, which correspond to 1$\sigma$ and 2$\sigma$.}
\label{fig:vel}
\end{figure*}
%%%%%%%%%%%%%%%%%%%%%%%%%%%%%%%%%%%%%%%%%%%%%%%%%%%%%%%%%%%%%%%%%%%%%%%

An interesting comparison includes the WNH stars, which are usually
thought to be approaching the end of core-H burning main sequence
evolution in the most luminous stars (see, e.g.,
\citealt{ms79,crowther95,drissen95,dekoter97,sc08}).  At this point in
their evolution near the end of the main sequence, their luminosities
have increased and their stellar mass has decreased somewhat due to
mass loss, so that their proximity to the Eddington limit causes
strong winds \citep{graefener11,sc08}.  One would infer that these
stars should have environments the same as LBV1 and fairly well
associated with early O stars.  We see that indeed, WNH stars are
closely associated with early O stars.  However, WNH stars have a
distribution very unlike the classical LBVs, which should be their
immediate descendants (the expected age difference between WNH and
LBV1 is less than 1 Myr).  Instead, the very luminous LBVs are more
closely associated with late O stars and WC stars.  Overall, this
comparison to the locations of early O stars reinforces the anomalous
locations of LBVs found by ST15, and directly contradicts claims made
by H16.

Whether or not runaways are required to explain the dispersal of LBVs
--- or if instead their apparent dispersal on the sky can be explained
by rejuvenation by gaining mass or merging in binaries --- remains an
open question in need of additional quantitative investigations with
population synthesis models or other techniques.  ST15 suggested that
kicks may help explain the isolation, but they did not claim that such
runaway motion is required.  Whether or not there is any observational
indication of runaway LBVs from kinematics is worth investigating as
well.  Nevertheless, it is clear that the apparent dispersal on the
sky of LBVs is certainly not consistent with expectations of the
traditional view in single-star evolution \citep{hd94} because they
clearly do not associate with stars of similar expected initial mass,
even if they tolerate some moderately massive stars in their vicinity.

\section{Radial velocities}

In addition to the suggestion to subdivide the LBV sample as noted
above, H16 also claimed that none of the LBVs in M31/M33 have high
velocities consistent with being runaway stars, and that this supports
the view of LBVs as single stars.  H16 listed values for the observed
radial velocity $V_r(Obs.)$ of LBVs and the expected velocity
$V_r(Exp.)$ for M31 and M33 rotation curves in their Table 1.  They
did not present any analysis of these values, but claimed that they
contradict runaway motion.

An important point, though, is that the expected runaway speeds are
rather small (or order 10 km s$^{-1}$ or even less) in the
evolutionary scenario discussed for LBVs (see Sect.\ 2). Because the
expected speeds are so slow, it is not at all clear that available
radial velocity information for LBVs has the precision required to
rule out this motion.  In fact, a quantitative analysis shows that it
does not.

Figure~\ref{fig:vel} shows cumulative distributions of the residual
velocities for LBVs in M31 and M33.  These are the absolute value of
the residual $V_r(Obs.) - V_r(Exp.)$ for the LBVs listed in Table 1 of
H16 (in dark blue) compared to the same quantity for RSGs of ``rank
1'' (i.e. highly likely membership) from \citet{drout09} for M31 and
\citet{drout12} for M33 (RSGs shown in red).  In both cases, the LBVs
appear skewed to higher velocities by a small amount, although the
small number of LBVs do not allow a statistically significant
difference to be determined.  Note that most of the velocities seen
here probably result from uncertainty in $X$ and $R$ for individual
sources as compared to the model rotation curve for each galaxy (see
\citealt{drout09,drout12}).  Since M31 has a higher inclination and
higher intrinsic rotation speed, the residuals are naturally higher.
This makes it difficult to say anything about relatively slow runaways
from the observational data for M31, while M33 is more favorable.

Also shown in Figure~\ref{fig:vel} are empirical models for an
expected distribution of velocities for stars that receive an extra
kick.  These are made from the observed RSG distribution for each
galaxy, plus a randomized velocity from 0 km s$^{-1}$ to the nominal
kick velocity $V_k$. This is an idealized case where all LBVs have had
their companion explode already, and all kicks have the same velocity
$V_k$ but are isotropic, resulting in an equal probability at any
radial velocity between 0 km s$^{-1}$ and $V_k$ (in other words, the
line profile from a thin expanding spherical shell of finite thickness
is flat-topped).  This is certainly an oversimplification, but it is
conservative, and is suitable for the purpose of showing that it is
difficult to rule out some runaways.  When such extra motions are
added to the RSG distribution after randomization, it results in the
light blue distributions shown in Figure~\ref{fig:vel}.  Two model
distributions are shown for each galaxy; one is the case with the
highest p-value in a Kolmogorov-Smirnov (KS) test, and the other has a
p-value equal to that of the RSG population (in other words, an equal
probability of having either no additional kick or this kick speed,
respectively).

For M31, the data allow rather high runaway speeds for the full
distribution.  The highest p-value (84 per cent) is for $V_k$=20 km
s$^{-1}$, and additional kicks as high as $V_k$=50 km s$^{-1}$ have
equal probably to zero kick (25 per cent).  For M31, the data appear
to be fully consistent with the hypothesis that all LBVs could have
received motion larger than typical speeds expected in the mass-gainer
scenario for LBVs (ST15).

For M33, the data are somewhat more constraining and higher speeds may
be less likely.  The highest probability is for $V_k$=5 km s$^{-1}$
(99 per cent), and additional kicks as high as $V_k$=8 km s$^{-1}$
have equal probably to no kicks (93 per cent).  As noted earlier,
relatively small kicks of only a few to 10 km s$^{-1}$ are expected
from mass gainers in binary evolution \citep{demink14,eldridge11}, so
the velocities for LBVs in M33 still permit this scenario for all
LBVs.  Strictly speaking, even larger speeds cannot be ruled out.  The
two additional magenta curves in Figure~\ref{fig:vel} (right) show
simulated distributions for kicks of 20 (p=32 per cent or 1$\sigma$)
and 35 km s$^{-1}$ (p=5 per cent or 2$\sigma$).  These allowed speeds
would increase if we admit that not necessarily every LBV has already
had its companion explode.

One must bear in mind, of course, that since most massive stars are
born in interacting binary systems \citep{sana12,dk13}, the observed
populations of RSGs in M31 and M33 (which served as a reference
sample) {\it already contain a substantial fraction of stars that have
  received recoil motion} from their companion's SN.  Thus, some
runaways were already included by nature in the observed RSG
distribution, so the true kicks allowed by the data are even larger
than the numbers quoted above.

Admittedly, it would be strange if the LBVs in M31 received
systematically higher kicks than the LBVs in M33, as might be implied
by their distributions.  As noted above, however, most of the observed
residual velocities can be attributed to uncertainty in deriving
$V_r(Exp.)$ for a given star, which is larger for M31.  Residual
velocities are also strongly affected by errors in determinig
$V_r(Obs.)$, which is difficult for LBVs that have strong and variable
winds.  The main conclusion of this exercise is that the available
data do not have the precision required to discriminate between the
hypotheses that LBVs are or are not runaways, aside from ruling out
very high speeds of order 100 km s$^{-1}$ for most of the sample, much
larger than expected from binary evolutionary scenarios.

H16 also mentioned the velocities of LBVs in the Magellanic Clouds and
claimed that their kinematics suggest that they are single stars and
not runaways.  However, H16 only listed the velocities and compared
them to the average systemic velocity of the LMC without a
quantitative analysis.  The discussion of Magellanic Cloud LBV
kinematics is postponed to a later paper because this investigation
requires a different type of analysis than for M31/M33, and requires
new data.  However, it is noteworthy that the range of differential
velocities listed by H16 spans 10s of km s$^{-1}$, and appears at
first glance to be consistent with the relatively low-speed runaways
expected in the mass gainer scenario (ST15).  

One LBV is particularly interesting. R71 --- the most isolated of the
LBVs in the LMC, located more than 300 pc from any O-type star --- has
a radial velocity that is offset by $-$71 km s$^{-1}$ relative to the
systemic velocity of the LMC (according to Table 3 in H16).  At that
speed R71 could easily reach its observed isolation in a few Myr.

\section{Summary}

In conclusion, a quantitative look at the data of LBVs and their
environments shows that even if one adopts the selective criteria
advocated by H16, the results do not support their claims.  When only
a few objects from the tail of a distribution are extracted, there
remains no statistical power to discriminate between that sample and
the remainder.  Moreover, the analysis above shows that even if those
selections are permitted, the interpretation arrived at by H16 was
incorrect, because they did not consider the quantitative implications
of the comparison stars.  Namely, one finds that the most luminous
confirmed LBVs have environments similar to late O-type stars, which
have median ages about twice as long as the presumed ages of those
LBVs in a single-star scenario.  Similarly, the lower-luminosity
confirmed LBVs have a distribution similar to RSGs, the bulk of which
have initial masses (10$-$15 $M_{\odot}$), less than half that of the
low-luminosity LBVs (25$-$40 $M_{\odot}$). This discrepancy rules out
the traditional single-star view of LBVs, and requires instead that
they are massive blue stragglers.

A statistical test of the confirmed and candidate LBVs shows no
statistical difference between their environments, contrary to the
central claim that motivated H16's reanalysis.  Thus, it
is not clear that it is appropriate to separate them.

Even if one does separate the confirmed and candidate LBVs, and if one
further separates the low and high luminosity group of LBVs, the
central results of ST15 remain the same -- that LBVs are more isolated
from O-type stars than they should be in the traditional single star
view of stellar evolution.  Rejuvenation by mass transfer and mergers,
and possibly runaway motion from a companion's SN, are required to
explain their isolation.  Available kinematics of LBVs do not argue
against SN-induced runaways, mostly because the expected motion is
slow and the precision of available data cannot clearly discriminate.
It will be interesting to see if there are other indications that LBVs
do not have anomalous motion; if they do not, then this will point to
rejuvenation in binary evolution (i.e. massive blue stragglers) as the
main explanation for their isolated environments.

%\acknowledgments \footnotesize
\smallskip\smallskip\smallskip\smallskip
\noindent {\bf ACKNOWLEDGMENTS}
\smallskip
\footnotesize

I thank an anonymous referee for helpful suggestions.  Support was
provided by NSF awards AST-1312221 and AST-1515559, and by the
National Aeronautics and Space Administration (NASA) through HST grant
AR-14316 from the Space Telescope Science Institute, which is operated
by AURA, Inc., under NASA contract NAS5-26555. This research has made
use of the SIMBAD database, operated at CDS, Strasbourg, France.

% REFERENCES

\end{document}